\documentclass[12pt,preprint]{aastex61}
\usepackage{epsfig,amsfonts,amsmath,amssymb,graphicx,morefloats,appendix,tensor}
\usepackage{color}
\usepackage{natbib}
\begin{document}

\title{Resolved Millimeter Observations of the HR 8799 Debris Disk}

\author{David J. Wilner}
\affiliation{Harvard-Smithsonian Center for Astrophysics, 60 Garden Street, 
       Cambridge, MA 02138, USA}

\author{Meredith A. MacGregor}
\altaffiliation{NSF Postdoctoral Fellow} 
\affiliation{Harvard-Smithsonian Center for Astrophysics, 60 Garden Street, 
       Cambridge, MA 02138, USA}  
\affiliation{Department of Terrestrial Magnetism, Carnegie Institution for 
Science, 5241 Broad Branch Road, Washington, DC 20015, USA} 

\author{Sean M. Andrews}
\affiliation{Harvard-Smithsonian Center for Astrophysics, 60 Garden Street, 
       Cambridge, MA 02138, USA}

\author{A. Meredith Hughes}
\affiliation{Department of Astronomy, Van Vleck Observatory, Wesleyan 
University, 96 Foss Hill Drive, Middletown, CT 06459, USA}

\author{Brenda Matthews}
\affiliation{National Research Council of Canada, Herzberg Astronomy and 
Astrophysics Programs, 5071 West Saanich Road, Victoria, BC, V9E 2E7, Canada}

\author{Kate Su}
\affiliation{Steward Observatory, University of Arizona, Tucson, AZ 85721, USA}

\email{dwilner@cfa.harvard.edu}

\begin{abstract}
We present 1.3~millimeter observations of the debris disk surrounding 
the HR~8799 multi-planet system from the Submillimeter Array to 
complement archival ALMA observations that spatially filtered away the bulk 
of the emission. The image morphology at $3\farcs8$ (150~AU) resolution 
indicates an optically thin circumstellar belt, which 
we associate with a population of dust-producing planetesimals within the
debris disk.
The interferometric visibilities are fit well by an axisymmetric radial 
power-law model characterized by a broad width, $\Delta R/R\gtrsim 1$.
The belt inclination and orientation parameters are consistent with the 
planet orbital parameters within the mutual uncertainties. 
The models constrain the radial location of the inner edge of the belt 
to $R_\text{in}= 104_{-12}^{+8}$~AU. In a simple scenario where the chaotic 
zone of the outermost planet~b truncates the planetesimal distribution, 
this inner edge location translates into a constraint on 
the planet~b mass of $M_\text{pl} = 5.8_{-3.1}^{+7.9}$~M$_{\rm Jup}$. 
This mass estimate is consistent with infrared observations of the planet 
luminosity and standard hot-start evolutionary models, with the uncertainties 
allowing for a range of initial conditions.  We also present new 
9~millimeter observations of the debris disk from the Very Large Array and 
determine a millimeter spectral index of $2.41\pm0.17$. This value is typical 
of debris disks and indicates a power-law index of the grain size distribution 
$q=3.27\pm0.10$, close to predictions for a classical collisional cascade. 
\end{abstract}


\section{Introduction}
\label{sec:intro}

The young \citep[30 Myr,][]{malo13} and nearby \citep[39.4 pc,][]{van07}
A-type star HR~8799 is the host of the first (and so far only) directly 
imaged multiple planet system. Near-infrared images show four companions 
with projected separations of 14, 24, 38, and 68 AU \citep{marois08,marois10} 
whose orbital motions have been tracked over a decade 
\citep[e.g. see the compilation by][]{bowler16}.  
Comparison of infrared photometry with standard (``hot-start'') 
evolutionary models suggest these companions have masses in the range of
$5-10$~M$_{\rm Jup}$ \citep{marley12}. This inference is consistent with 
calculations that imply planet masses $<10$~M$_{\rm Jup}$ for dynamical 
stability at the age of the system, which also could be bolstered by 
a 1:2:4:8 mean motion resonance configuration 
\citep[e.g.][]{gozdziewski14,pueyo15,maire15}. 
Analysis of self-consistent and homogeneous astrometric measurements
indicate that the planet orbits have low eccentricity and are consistent 
with coplanarity \citep{konopacky16}. These relatively bright and 
wide-separation super-Jovian planets are also a favorite target for 
spectroscopic observations aimed at planetary atmosphere characterization 
\citep{barman11,konopacky13,ingraham14,barman15,bonnefoy16,zurlo16}.

The HR~8799 system hosts, in addition to planetary mass companions, a 
dusty debris disk that was first detected by {\em IRAS} \citep{sadakane86}.
The debris disk has been investigated in detail at wavelengths from 
$24-850~\mu$m using observations from {\em Spitzer} \citep{su09}, 
{\em Herschel} \citep{matthews14}, 
and the JCMT \citep{williams06,holland17}.
These multi-wavelength observations show that the debris consists of a 
warm (T$\sim150$~K) inner belt and a cold (T$\sim35$~K) outer belt that 
bracket the orbits of the directly imaged planets, plus an extended halo of 
small grains that is detected out to radii beyond $1000$~AU. 
The 850~$\mu$m photometry marks the HR~8799 debris disk as one of the 
most massive known, at an estimated mass of $\sim0.1$~M$_{\rm Earth}$.
Modeling of resolved far-infrared images suggests the presence of a population 
of colliding planetesimals underlying the cold belt that extends from 
about $100$ to $300$~AU, albeit with significant uncertainty on the
locations of these boundaries on account of insufficient angular resolution 
for the inner edge and confusion with the extended halo. 
The relatively large radial extent 
of this belt coupled with the young age of the system supports 
``planet-stirring'' models to produce the collisional debris \citep{moor15}.

Millimeter emission selectively reveals the large dust grains less
affected by radiative forces that therefore trace best the distribution 
of dust-producing planetesimals within debris disks \citep{wyatt06,wilner11}. 
Single dish observations of HR~8799 from the CSO at $350~\mu$m 
($9''$ beam) resolved emission from the cold belt around the star, with 
a tentative offset asymmetry \citep{patience11}. This asymmetry was not 
confirmed by observations with higher signal-to-noise ratio from 
the JCMT at 450~$\mu$m and 850~$\mu$m ($8''$ and $13''$ beam, 
respectively) \citep{holland17}. Millimeter imaging of the disk at much 
higher angular resolution has proven challenging on account of its
low surface brightness.  \citet{hughes11} made the first millimeter 
interferometric detection of the disk, using the Submillimeter Array (SMA) 
at 870~$\mu$m, consistent with the presence of a broad belt 
($\Delta R/R > 1$) of inner radius $\sim150$~AU. 
Recently, \cite{booth16} used the Atacama Large Millimeter/submillimeter 
Array (ALMA) to image the HR~8799 system at 
1.3~mm at much higher sensitivity, although these observations recovered 
only a small fraction of the total disk emission. Analysis of these 
ALMA data by an unconventional fitting of belt models to dirty images 
placed apparently strong constraints on the structure. In particular, 
their analysis determined an inner edge too far out to be truncated by 
the outermost planet~b and raised the possibility of the presence of an 
additional (and unseen) outer planet. 

To better determine the properties of the cold belt of planetesimals in 
the HR 8799 system, we used the SMA to obtain new observations at 
1.3~millimeters wavelength that are sensitive to larger angular scales 
than the previous millimeter interferometer studies by \cite{hughes11} and 
\citet{booth16}. We also obtained new observations at 9~millimeters 
wavelength with the Karl G. Jansky Very Large Array (VLA), to constrain the 
spectral index of the disk emission. In \S2, we describe the details of these 
1.3~millimeter and 9~millimeter observations. In \S3, we present the 
1.3~millimeter results in the form of images and deprojected visibility 
functions, and provide a quantitative analysis of these data using parametric 
model fits in an MCMC framework. In \S4 we discuss implications of the new 
millimeter results for the system geometry, the mass of planet b, and the 
grain size distribution.  In \S5, we summarize the main conclusions.

\section{Observations}
\label{sec:obs}

\subsection{Submillimeter Array}
We observed the HR~8799 system with the SMA on Mauna Kea, Hawaii, 
at a wavelength near 1.3~millimeters in the subcompact configuration
that provides projected baselines as short as 6~meters. 
Table~\ref{tab:sma} provides a log of these observations, including 
the observing dates, weather conditions, number of operational antennas, 
and the useable hour angle range. 
In this close-packed antenna configuration, the lower elevation limit 
for observations is $33\degr$, to prevent antenna collisions.  
The phase center was set at $\alpha=23^\text{h}07^\text{m}28.72$, 
$\delta=+21\degr08\arcmin03\farcs3$ (J2000),
i.e. the J2000 star position uncorrected for interim proper motion.
Two units of the SWARM digital correlator, still under construction
at the time, were available; together these spanned bands $\pm(4-12)$~GHz 
from the LO frequency of 225.5 GHz, with uniform channel spacing 140~kHz 
($\sim$0.18~km~s$^{-1}$). This setup provided a total of 16~GHz of continuum 
bandwidth, as well as simultaneous coverage of the 
$^{12}$CO J=2-1 (230.53800 GHz) 
and $^{13}$CO J=2-1 (220.39868 GHz) 
spectral lines in the upper and lower sidebands, respectively.
The primary beam FWHM size of the 6-meter diameter array antennas 
of $55\arcsec (\nu/230~{\rm GHz})$ set the useable field of view. 

The basic observing sequence consisted of a loop of 2 minutes each 
on the quasars 3C454.3 ($\sim13$~Jy, $6\degr$ away) 
and J2232+117 ($\sim5$~Jy, $12\degr$ away) and 10 minutes 
on the target source HR~8799.  Passband calibration was obtained
with observations of the strong sources 3C454.3 and Uranus.
The absolute flux scale was set using observations of Uranus
obtained in each track, with 10\% overall accuracy. 
All of the basic calibration was performed using standard procedures
in the MIR software package. Continuum visibilities were output in 
30 second scans spanning 4 GHz widths, centered at 215.5, 219.5, 
231.5, and 235.5 GHz.
Fourier inversion for continuum and spectral line imaging and {\tt clean} 
deconvolution were done in the {\tt CASA} software package (version 4.7.2).

\begin{deluxetable}{l c c c c}
\tablecolumns{5}
\tabcolsep0.06in\footnotesize
\tabletypesize{\small}
\tablewidth{0pt}
\tablecaption{\label{tab:sma}
SMA 1.3 Millimeter Observations of HR 8799} 
\tablehead{
\colhead{Observation} & \colhead{\# of } & \colhead{Projected} & 
\colhead{$\tau_{atm}$} & \colhead{H.A.} \\
\colhead{Date} & \colhead{Antennas} & \colhead{Baselines (m)} & 
\colhead{225~GHz} & \colhead{range}
}
\startdata
2016 Aug 17 & 7 & $7-69$ & 0.08 & $-3.9,+3.7$ \\
2016 Aug 27 & 8 & $6-69$ & 0.08 & $-4.0,+3.8$ \\
2016 Aug 29 & 8 & $6-69$ & 0.10 & $-1.7,+3.7$ \\
2016 Aug 30 & 8 & $6-69$ & 0.07 & $-0.8,+3.7$ \\
2016 Sep 06 & 8 & $6-69$ & 0.09 & $-3.8,+3.7$ \\
2016 Sep 21 & 7 & $6-45$ & 0.08 & $-3.6,+3.6$ \\ 
\enddata
\end{deluxetable}

\subsection{Atacama Large Millimeter/submillimeter Array}

We retrieved observations of HR 8799 from the ALMA archive that were
made in Band~6 in 2015 January. Details of these observations are described 
in \citet{booth16}. We calibrated the five individual scheduling block 
executions using the associated \texttt{CASA} 4.3.1 reduction scripts. 
In brief, the correlator setup consisted of four spectral windows with 
width 2 GHz and channel spacing 16 MHz, centered at 216, 218, 231 and 233 GHz, 
respectively.  
The observations included baseline lengths from 15 to 349 meters. 
The primary beam FWHM size of the 12-meter diameter array antennas 
of $27\arcsec (\nu/230~{\rm GHz})$ set the useable field of view. 
We used the \texttt{CASA} task {\tt statwt} to homogenize the visibility 
weights among scheduling blocks and note that the execution on 
2015 January 03 obtained in the best weather conditions dominates 
the sensitivity budget.
Continuum visibilities were output for analysis after flagging the
channels with CO line emission, averaging in time to 30 seconds, and 
averaging in frequency to 2 GHz bandwidth.  Imaging was carried out
with the {\tt clean} task in {\tt CASA} (version 4.7.2).

\subsection{Very Large Array}
We observed the HR~8799 system with the VLA at 9~mm wavelength in the 
most compact D configuration.  Observations with the 27 array antennas were 
executed in 2 hour scheduling blocks on 4 dates in early 2017: 
Feb 16, Mar 04, 11, and 13. The baselines ranged from 0.04 to 1.03 km. 
The weather conditions were very good, with phase noise measurements from 
the Atmospheric Phase Interferometer (on a 300 m baseline at 11.7~GHz) 
ranging from 3.3 to 7.4$\degr$.  The phase center was set to the same 
position as for the SMA observations. 
The correlator was used to obtain the maximum continuum bandwidth of 8 GHz, 
comprised of 4 bands centered at 30, 32, 34, and 36 GHz, for two polarizations.
The primary beam FWHM size of the 25-meter diameter array antennas is
approximately $85\arcsec (\nu/ 30~\rm GHz)$.
The observing sequence interleaved 1 minute observations of the 
complex gain calibrator 3C454.3 with 5 minute observations on HR~8799.  
Passband calibration was obtained with observations of 3C454.3. 
The absolute flux scale was set with observations of the standard calibrator 
3C48 in each execution, with an estimated accuracy of 10\%.  The basic 
calibration was done using VLA pipeline processing in the {\tt CASA} package 
(version 4.7.2) followed by imaging with {\tt clean} task.

\section{Results and Analysis}
\label{sec:results}

\begin{figure}[h]
\begin{center}
\includegraphics[scale=0.95,angle=0]{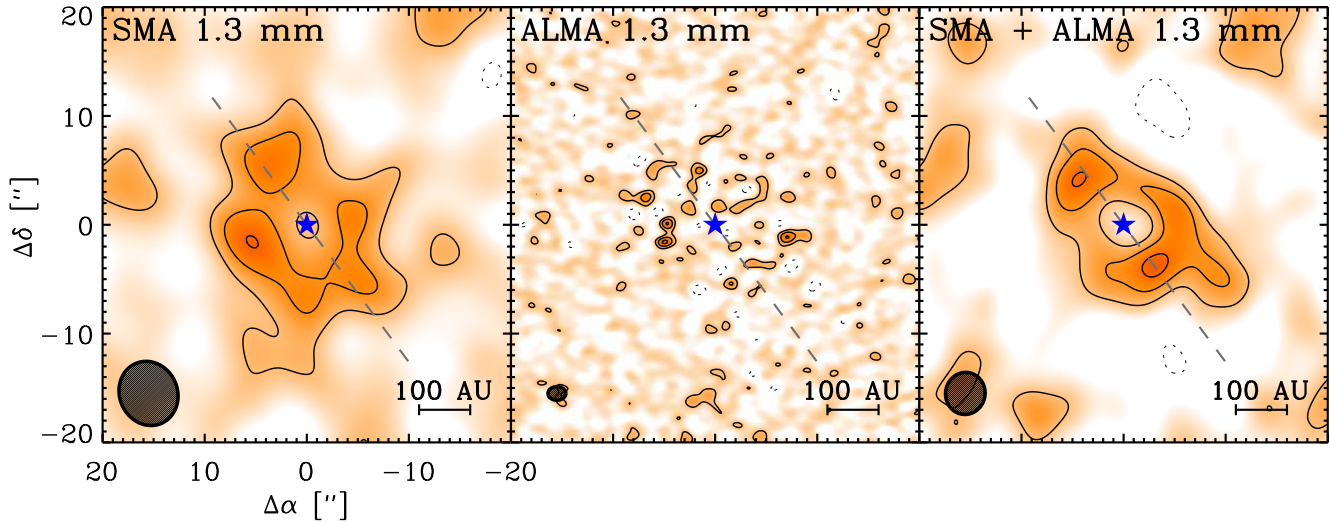}
\figcaption{
Images of 1.3~mm continuum emission from the HR 8799 debris disk 
from the SMA (left), ALMA (center), and the SMA and ALMA combined (right). 
All of these images were made with natural weighting, and those including 
SMA data make use of a $4\arcsec$ FWHM Gaussian taper. Contour levels are in 
steps 
of $2\times$ the measured rms values are 180~$\mu$Jy/beam, 16~$\mu$Jy/beam, 
and 30~$\mu$Jy/beam for the three images, respectively.  
The beam sizes are shown by the ellipses in the lower left corner of each 
image and are $6\farcs1\times5\farcs6$, $1\farcs7\times1\farcs2$, and 
$3\farcs8\times3\farcs7$, respectively.
The blue star symbol indicates the stellar position, and the dashed gray line 
shows the best-fit belt position angle (see \S\ref{sec:model}). 
}
\end{center}
\label{fig:cont}
\end{figure}

\begin{figure}[h]
\begin{center}
\includegraphics[scale=1.00,angle=0]{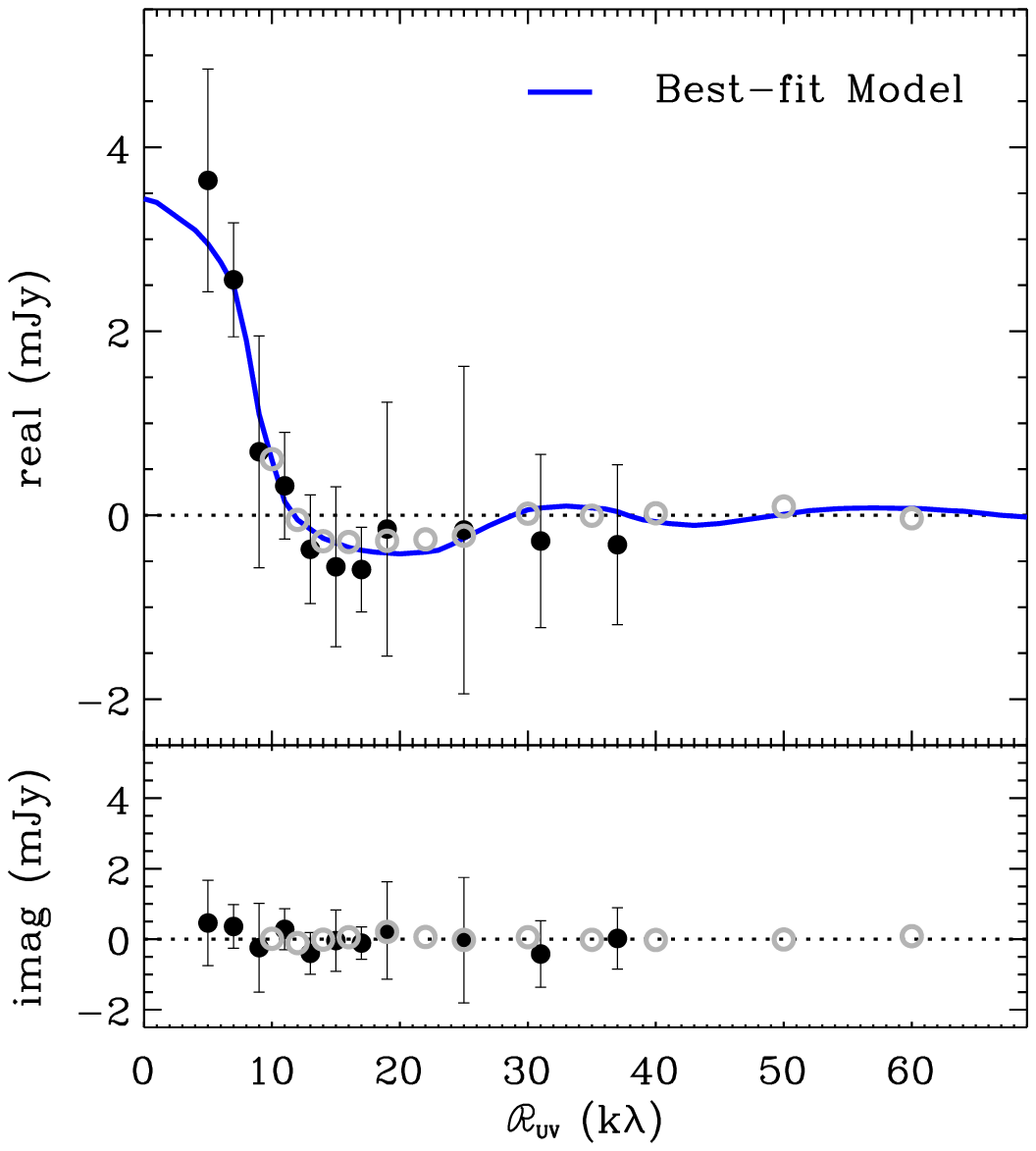}
\figcaption{
The real and imaginary parts of the 1.3~mm visibilities, averaged in bins of 
deprojected $(u,v)$ distance from the star, for data from the SMA 
(black filled circles) and ALMA (gray open circles). 
The solid blue line depicts the best-fit power law axisymmetric belt 
model to the combined dataset (see \S\ref{sec:model}).  The imaginaries
are consistent with zero as is expected for an axisymmetric belt.
}
\end{center}
\label{fig:vis}
\end{figure}

\subsection{Continuum Emission}
\label{subsec:continuum}
Figure~\ref{fig:cont} shows 1.3~mm continuum images of HR~8799 from 
the SMA (left), ALMA (middle), and the combination of SMA and ALMA (right). 
These images were obtained with natural weighting, and those including 
SMA data use a $4\arcsec$ FWHM Gaussian taper to improve 
surface brightness sensitivity at the expense of angular resolution.
The ALMA image is comparable to Figure~1 of \citet{booth16}, with a 
beam size of $\sim1\farcs5$, and it shows a noisy ring of emission 
surrounding the star.
For the SMA image, the beam size is $\sim4\times$ larger in each dimension
than for the ALMA image, and the noise level is an order of magnitude higher, 
but the sensitivity to larger angular scales provided by the shorter baselines 
enables a much improved representation of the radially extended belt of 
emission around the star. The combination of SMA and ALMA data results in an 
even better image of this circumstellar structure. In the SMA and ALMA 
combined image shown with a $\sim3\farcs8$ beam, the two peaks visible on 
either side of the star are the signatures of limb brightening along the 
major axis of an inclined and optically thin emission belt. 

Figure~\ref{fig:vis} shows the deprojected visibility function for the 
emission from the SMA and ALMA, obtained by averaging the real and imaginary 
parts of
of the complex visibilities in concentric annular bins, adopting the belt 
geometry derived in \S\ref{sec:model}. This view of the visibilities 
provides some insight into the image properties in Figure~\ref{fig:cont}. 
The shorter baselines from the SMA capture the steep rise in visibility 
amplitude at low spatial frequencies that are missed by ALMA, and thus 
sample better the emission responsible for the extended disk morphology.
The visibilities from the two telescopes are consistent where they overlap 
at higher spatial frequencies that sample the small scale structure.  
The imaginary parts of visibilities are all consistent with zero, as expected 
for an axisymmetric structure.

\begin{figure}[ht!]
\begin{center}
\includegraphics[scale=0.50,angle=0]{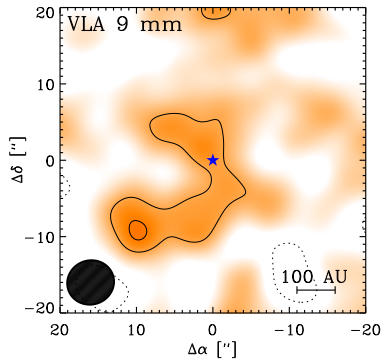}
\figcaption{
Image of the 9~millimeter continuum emission from the HR 8799 debris disk
from the VLA obtained with natural weighting and a $5\arcsec$ FWHM Gaussian
taper.  Contour levels are $\pm3,5\times$ the measured rms of 
4.3~$\mu$Jy/beam. The $6\farcs2\times5\farcs8$ beam size is indicated by the 
ellipses in the lower left corner and the stellar position by the blue star symbol.
}
\end{center}
\label{fig:vla}
\end{figure}

Figure~\ref{fig:vla} shows a VLA 9~millimeter continuum image of the HR~8799 
system, obtained with natural weighting and a $5\arcsec$ FWHM Gaussian taper, 
which gives a beam size of $6\farcs2 \times 5\farcs8$ and rms noise of
4.3~$\mu$Jy/beam. 
Significant emission is detected from an extended region centered on the star, 
but with a low peak signal-to-noise ratio of only $3-4$ per beam. No clear 
disk or belt morphology is evident in the emission at this sensitivity level. 
The emission peak near $(+10\arcsec,-10\arcsec)$ in the image could be the
result of a noise fluctuation, or perhaps a faint and unrelated background 
source.  

\subsection{Line Emission}
\label{sec:line}
\begin{figure}[ht!]
\begin{center}
\includegraphics[scale=0.87,angle=0]{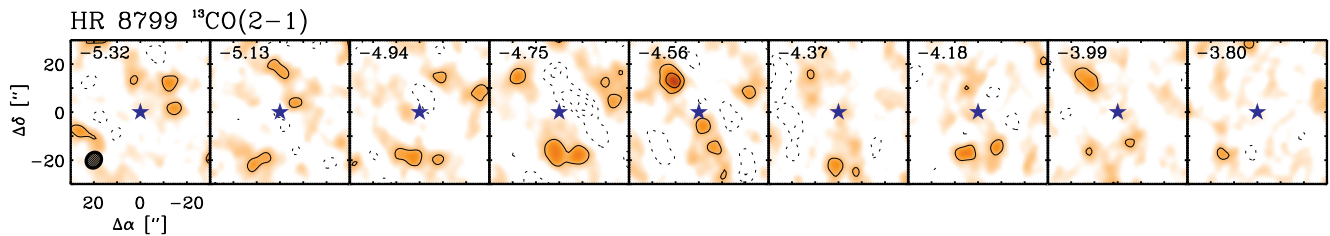}
\figcaption{
SMA channel maps of $^{13}$CO $J=2-1$ emission for HR 8799. 
For all panel, the contours are in steps of $\pm2\times$ the rms noise level 
of 100 mJy/beam.  The ellipse in the lower left corner of the leftmost panel 
indicates the $7\farcs0\times6\farcs1$ synthesized beam size resulting
from natural weighting and a $4\arcsec$ Gaussian taper.  
The LSR velocities are labeled in the upper left corner of each panel.
}
\end{center}
\label{fig:13CO}
\end{figure}

Emission from the $^{12}$CO and $^{13}$CO J=2-1 lines were clearly detected 
by the SMA observations over the narrow $v_{\rm LSR}$ range $-6$ to $-4$ 
km~s$^{-1}$, close to the stellar velocity of $-4.6$~km~s$^{-1}$ in this 
reference frame. 
Single dish observations of the $^{12}$CO J=3-2 line with a $15''$ beam 
show extended emission, concentrated in a filament that extends from the 
north-northeast of HR~8799 to the south-southwest, most likely affiliated with 
the nearby MBM 53-55 complex of high latitude clouds \citep{williams06,su09}. 
The submillimeter dust continuum emission observed by {\em Herschel} also 
shows traces of this large scale feature \citep{matthews14}.
Like previous attempts to use interferometers to image the 
molecular emission in this region, the extended structure evident in lower 
angular resolution single dish maps \citep{su09} remains poorly sampled 
and highly problematic to reconstruct, even with the lower spatial 
frequency information newly obtained with the SMA.  Regardless of the choice of 
visibility weighting scheme and {\tt clean} deconvolution parameters, all of 
our attempts at imaging the line emission show pronounced artifacts.  
Emission from the $^{13}$CO J=2-1 line, with substantially lower
optical depth than the main isotopolgue, offers the best chance to reveal
structure associated with the HR~8799 system.
Figure~\ref{fig:13CO} shows a set of $^{13}$CO channel maps 
obtained with natural weighting and a $4\arcsec$ FWHM Gaussian taper, 
resulting in a $\sim6\farcs5$ beam. Among the positive and negative 
corrugations indicative of missing flux, the most significant emission is 
located away from the star, and there is no clear evidence for any systematic 
motions that might be associated with rotation of gas in the debris disk. 
The linewidth is consistent with the typical turbulent broadening found 
in small molecular clouds. The apparent association in space and velocity 
between HR~8799 and this extended molecular emission is intriguing but 
remains mysterious.

\subsection{SMA/ALMA Disk Model Fits}
\label{sec:model}
To provide a quantitative characterization of the 1.3 millimeter emission 
from the HR 8799 disk, we make the simple assumption that the structure 
can be represented by an axisymmetric and geometrically thin belt, and we 
model the observations using the visibility fitting procedure described 
in \citet{macgregor13}. 
The surface brightness of the belt is assumed to be a radial power law, 
$I_\nu \propto r^x$, between an inner and outer radius, 
$R_\text{in}$ and $R_\text{out}$, respectively. The surface brightness 
power law index, $x$, encapsulates both the radial surface density and 
temperature profiles of the emitting grains. The total flux density of 
the belt is normalized to 
$F_\text{belt} = \int I_\nu d\Omega$.  We also fit for the belt geometry,
specifically an inclination, $i$ (where $0\degr$ is face-on), 
and a position angle, $PA$ (measured East from North).  
For comparison with observed visibilities, the model emission is 
multiplied by the appropriate primary beam response of each telescope. 
Since imaging with both {\em Herschel} \citep{matthews14} and JCMT
\citep{holland17} shows the presence of a compact background source to the 
northwest of the disk, we also include an unresolved point source at this 
location in the model, described by a total flux, $F_\text{pt}$, and an 
offset from the phase center of the observations, 
$\{\Delta\alpha_\text{pt},\Delta\delta_\text{pt}\}$. 
When modeling the ALMA data alone, this additional background source is 
omitted on account of primary beam attenuation at the offset location.

This simple two-dimensional parametric model for the disk emission is 
incorporated into a Markov Chain Monte Carlo (MCMC) framework. 
For each model image, we calculate synthetic visibilities using 
{\tt vis\textunderscore sample}\footnote{{\tt vis\textunderscore sample} 
is publicly available at \url{https://github.com/AstroChem/vis_sample} or in 
the Anaconda Cloud at \url{https://anaconda.org/rloomis/vis_sample}}, a python 
package for visibility sampling that improves on the sampling and interpolation of the {\tt  uvmodel} task in the Miriad software package, and determine a $\chi^2$ likelihood function, $\text{ln}\mathcal{L} = -\chi^2/2$, that incorporates the statistical weights of the visibilities.  We make use of the {\tt emcee} package \citep{foreman13} 
to sample the posterior probability distribution of the data conditioned on the model, and determine the best-fit parameter values and their uncertainties.  We explore the posterior with 50,000 iterations ($100$ walkers and $5,000$ steps each) and check that the Gelman-Rubin statistic \citep{gelman14} is $\hat{R}<1.1$ for all parameters to ensure convergence.

\begin{figure}[h]
\begin{center}
\includegraphics[scale=0.80,angle=0]{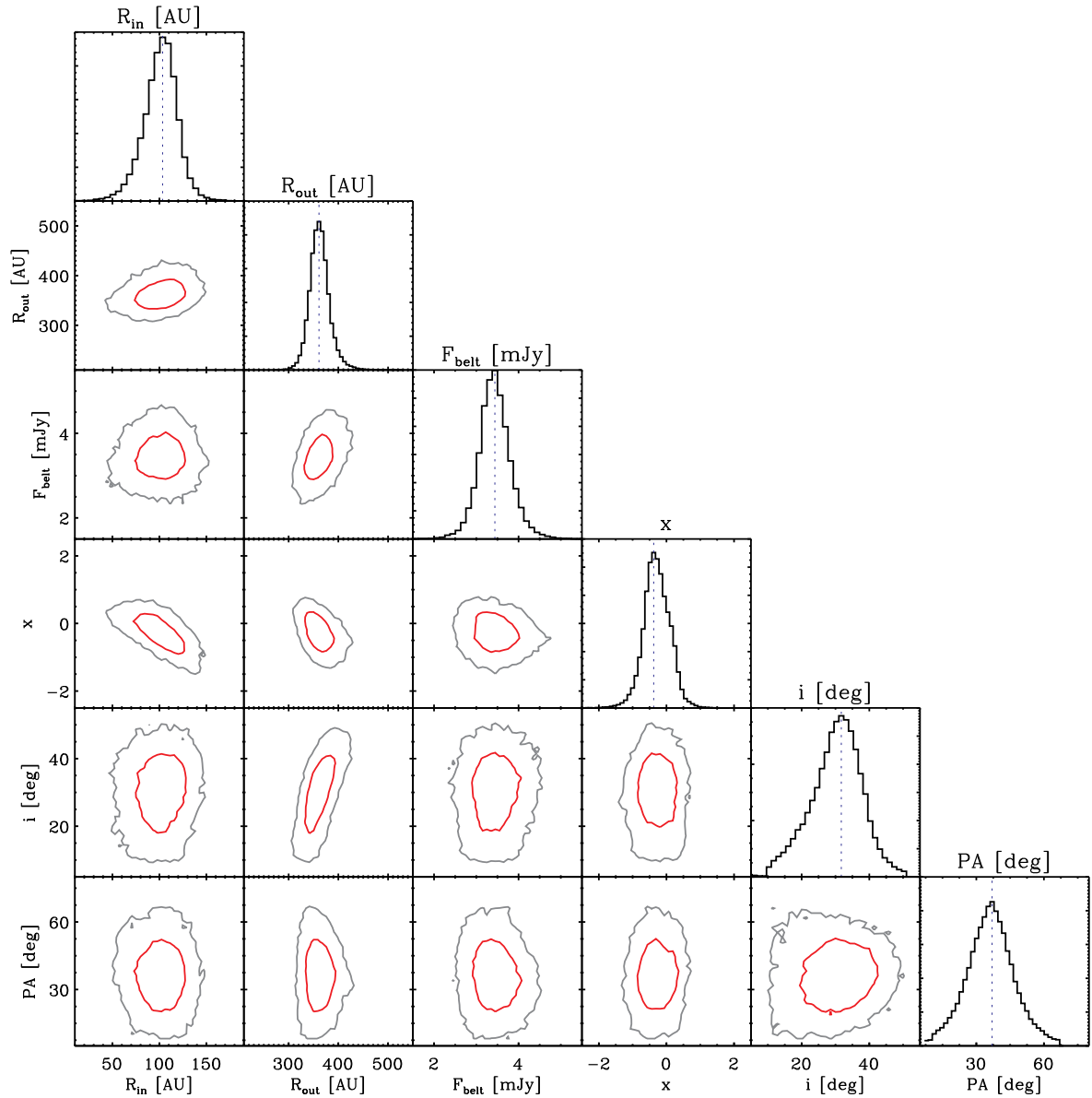}
\figcaption{
Corner plot summary of parameter values from the MCMC belt modeling
of the combined SMA and ALMA data. 
The panels on the diagonal show the 1-D histogram for each model parameter 
obtained by marginalizing over the other parameters, with a dashed 
vertical line to indicate the best-fit value. The off-diagonal panels 
show 2-D projections of the posterior probability distributions for each
pair of parameters, with contours to indicate $1\sigma$ (red) and $2\sigma$ 
(grey) regions.
}
\end{center}
\label{fig:cornerplot}
\end{figure}

We first fit the SMA observations and the ALMA observations separately, 
to examine the constraints provided by each dataset alone, and then 
fit the observations from both telescopes simultaneously, to examine 
the joint constraints. Table~\ref{tab:par} lists the best-fit parameters
from each of these fits, together with their 68\% uncertainties determined 
from the marginalized posterior probability distributions. 
For modeling the SMA observations alone, we adopt uniform priors for all 
parameters, in particular $0<R_\text{in}<R_\text{out}$, 
$F_\text{belt}>0$, and $-3<x<3$. When modeling the ALMA observations 
alone, informative constraints were not obtained for the outer radius 
or total flux parameters using these assumptions. In effect, the ALMA data 
allow the outer radius parameter to extend freely, with ever increasing 
total flux.  This most likely results because the ALMA primary beam is 
comparable in size to the HR~8799 emission region, and the ALMA 
observations provide very little data at spatial frequencies inside the 
first null of the visibility function to limit the total disk emission.
To obtain constraints on these parameters from the ALMA data alone, we must 
impose strong priors on both $R_\text{out}$ and $F_\text{tot}$.  As an example, 
we implemented a Gaussian prior on the total belt flux density, $3.0\leq F_\text{tot} \leq 4.0$~mJy, motivated by submillimeter observations \citep{matthews14} that predict a flux density of $\sim3.5$~mJy at 1.3~mm 
assuming a millimeter spectral index of $2.5$, typical of debris disks 
\citep{macgregor16}.  We note that
the fit to the SMA data alone yields a higher flux density of 
$5.13^{+0.34}_{-1.01}$~mJy, but still comparable within the uncertainties.
In addition, we implemented a logistic function $f(x) = 1/(1+e^{-k(x-x_0)})$ 
prior for the outer radius with $k=-10$ and $x_0 = 512$~AU, the primary beam diameter. 
With these assumptions, fitting the ALMA observations alone yields 
$F_\text{tot} = 3.5\pm0.5$~mJy and $R_\text{out} = 497.(+2.,-150)$~AU, 
effectively recovering the priors. 
We suspect that the $\sim10\%$ uncertainties on the disk parameters 
quoted by \citet{booth16} from an image-based analysis of the ALMA 
observations are underestimated, perhaps in part due to unaccounted 
covariance associated with large scale emission. By contrast, modeling the 
SMA and ALMA observations simultaneously does not require strong priors on 
the disk parameters to obtain meaningful constraints. 
Figure~\ref{fig:cornerplot} shows the posterior probability distributions 
for the model parameters that result from fitting the SMA and ALMA observations
together. For these combined fits, the small proper motion of HR~8799 
between SMA and ALMA observing epochs ($0\farcs20$) was ignored, given its 
insignificance at the low angular resolution of the SMA observations.

\begin{figure}[ht!]
\begin{center}
\includegraphics[scale=1.00,angle=0]{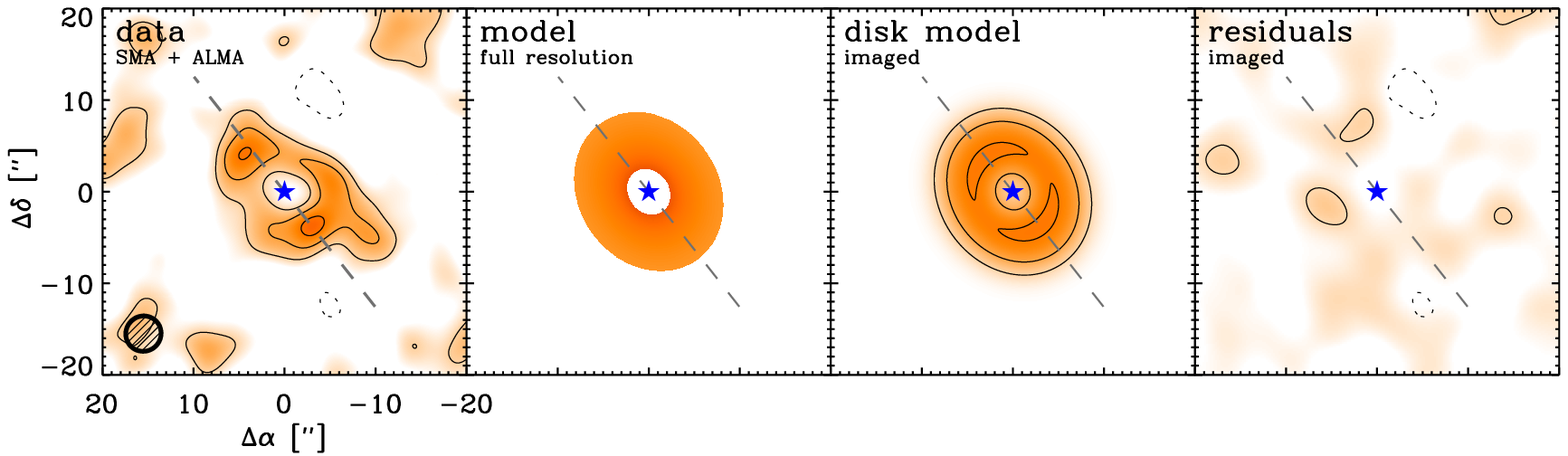}
\figcaption{
{\em (left)} The combined SMA and ALMA image of the HR 8799 debris disk.  
{\em (center left)} The best-fit model to the combined datasets displayed 
at full resolution (pixel size $0\farcs5=2$~AU at 39.4~pc).  
{\em (center right)}  The best-fit combined model convolved with the 
synthesized beam.  
{\em (right)}  The residuals obtained from imaging the best-fit model
subtracted from the data.
In all panels except the full resolution model, contours are in steps of 
$2\times30~\mu$Jy/beam, the rms noise level.  The blue star marks the 
stellar position, the dashed gray line indicates the best-fit position angle 
of the disk, and the dashed gray ellipse in the leftmost panel shows the 
$3\farcs8\times3\farcs7$ synthesized beam.
}
\end{center}
\label{fig:datmodres}
\end{figure}

Figure~\ref{fig:datmodres} shows the combined SMA and ALMA image from 
Figure~\ref{fig:cont} together with images of the best-fit model at full
resolution, the best-fit model imaged in the same way as the data, and 
an image made from the observed visibilities after subtracting the best-fit
model visibilities. This broad ($\Delta R/R\gtrsim 1$) inclined belt 
model reproduces very well the main features of the observations, including the
inner depression and limb-brightened regions in the northeast and southwest. 
The goodness-of-fit is demonstrated by the noise-like residual image. Notably, 
the imaged residuals reveal no significant azimuthal brightness asymmetries, 
as might arise from the presence of resonant clumps, or an offset of the star 
from the center of the disk. If any such asymmetries are present in the 
millimeter emission from the HR~8799 disk, observations with a significantly 
higher signal-to-noise ratio will be required to reveal them.
This outer belt of large fractional width joins a growing list other young 
debris disks with similar characteristics, e.g. HD~95086 \citep{su17}, that 
are much broader than the classical Kuiper Belt of our Solar System 
\citep[especially the low excitation ``kernel'' component discussed by][]{bannister16}.

\subsection{VLA 9~millimeter Flux Estimate}

In the VLA image, the disk is detected only at the $\sim3\sigma$ per beam
level, and we do not apply the same MCMC modeling technique to these data. 
For simplicity, we measure the flux density in {\tt CASA} within a circular 
aperture centered at the location of the disk, assuming the diameter of the 
disk determined from the combined modeling of the SMA and ALMA data.  
We determine the uncertainty by measuring the flux density in the same 
circular aperture at ten random locations in the image and taking the average. 
The measured total flux density of the disk from our VLA observations 
is $F_\text{9mm} = 32.6\pm9.9$~$\mu$Jy (correction for primary beam 
attenuation is negligible).  We adopt this image-based approach
instead of fitting to the millimeter visibilities due to the difficulty in 
selecting a function that adequately reproduces the complex structure of the 
observed surface brightness profile.

\begin{deluxetable}{l l c c c}
\tablecolumns{5}
\tabletypesize{\footnotesize}
\tablewidth{0pt}
\tablecaption{Best-fit Model Parameters \label{tab:par}}
\tablehead{
\colhead{Parameter} & 
\colhead{Description} & 
\colhead{SMA} &
\colhead{ALMA\tablenotemark{a}} &
\colhead{SMA+ALMA}
}
\startdata
$F_\text{belt}$ & Total belt flux density [mJy] & $5.13 (+0.34, -1.01)$ 
 & $3.50\pm0.50$ & $3.44 (+0.31, - 0.56)$ \\
$R_\text{in}$ & Belt inner edge [AU] & $106. (+19., -28.)$ 
 & $115. (+16., -17.)$ & $104. (+8., -12.)$ \\
$R_\text{out}$ & Belt outer edge [AU] & $370. (+42., -40.)$ 
 & $497. (+22., -150.)$ & $361. (+16., -18.)$ \\
$x$ & Surface brightness power law index & $-0.56 (+0.31, -0.83)$ 
 & $-0.64 (+0.40, -0.46)$ & $-0.38 (+0.47, -0.47)$ \\
$i$ & Disk inclination [$\degr$] & $35.6 (+5.6, -4.4)$  
 & $34.3 (+5.9, -6.9)$ & $32.8 (+5.6, -9.6)$  \\
$PA$ & Disk position angle [$\degr$] & $41.6 (+9.6, -7.8)$ 
 & $43.7 (+12.2, -12.4)$ & $35.6 (+9.4, -10.1)$ \\
\hline
$F_\text{pt}$ & background source flux density [mJy] & $0.47 (+0.11, -0.17)$  & \nodata & $0.45 (+0.12, -0.15)$  \\
$\Delta\alpha_\text{pt}$ & background source RA offset [$\arcsec$] & $-12.9 (+0.2, -0.9)$  & \nodata & $-12.2 (+0.8, -0.8)$  \\
$\Delta\delta_\text{pt}$ & background source DEC offset [$\arcsec$] & $17.8 (+0.4, -0.6)$ & \nodata & $17.5 (+0.9, -0.5)$ \\
\enddata
\tablenotetext{a}{Adopting strong priors on the total belt flux density
and belt outer edge parameters (see \S\ref{sec:model}).}
\end{deluxetable}

\section{Discussion}
\label{sec:discussion}
By modeling new and archival resolved millimeter observations, we have made 
an improved characterization of the planetesimal belt within the debris disk 
surrounding the HR~8799 planetary system. This broad emission belt spans from 
$\sim100-360$ AU, roughly consistent with previous modeling of the spectral 
energy distribution \citep{su09} and fitting far-infrared emission images
\citep{matthews14}.  The 1.3~millimeter interferometric dataset from SMA and 
ALMA together recovers the bulk of the disk flux and provides high angular 
resolution, and we used this combined dataset to fit for basic structural 
parameters. We next consider implications.  In particular, we examine 
the belt geometry constraints in the context of coplanarity and alignment of 
the planetary system (\S\ref{sec:geometry}), 
the relationship between the belt inner edge and the outermost planet~b 
(\S\ref{sec:planetmass}), 
ramifications of the radial surface density profile for 
collisional excitation mechanisms (\S\ref{sec:stirring}), 
and 
the millimeter spectral index connections to debris disk collisional models 
(\S\ref{sec:spectralindex}). 

\subsection{System Geometry}
\label{sec:geometry}
The model fits to the millimeter emission provide new estimates of the 
inclination and orientation of the planetesimal belt that can be 
compared directly to constraints on the orbital elements of the planets.
Formation of planets in a thin disk will generally lead to orbital coplanarity 
and alignment, unless upset by dynamical events. Since the orbital periods 
of the HR~8799 outer planets span centuries, only small arcs of their motions 
around the star have been measured in the years since discovery, leaving 
significant uncertainties in their orbital elements. Early stability 
calculations indicated departures from a face-on geometry \citep{fabrycky10}, 
and most analyses have favored orbits with inclinations of 10 to 30$\degr$ 
\citep[e.g.][]{currie12,esposito13,pueyo15,wertz17}. A recent analysis of 
carefully calibrated astrometric monitoring data that minimizes 
systematic errors generally indicate more highly inclined orbits, specifically 
$38\degr\pm7\degr$ for the outermost planet b, and no evidence for any 
measureable departures from coplanarity \citep{konopacky16}. 
Astroseismology suggests a higher inclination angle for the star,
$\gtrsim40\degr$, as well \citep{wright11}. These results are all compatible 
with the disk inclination of $32\fdg8^{+5.6}_{-9.6}$ derived from the 
combined SMA and ALMA data. 

There is some tension between the millimeter result and the lower 
(and more precise) disk inclination of $26\degr\pm3\degr$ determined from the 
ellipticity of the outer edge of far-infrared emission in {\em Herschel} 
images \citep{matthews14}. As noted by \citet{booth16}, the discrepancy 
between millimeter and far-infrared disk inclinations may have a physical 
origin in size-dependent dust dynamics, or in complexities in disk structure 
not captured by the simple belt models. Taking the millimeter emission as 
the best tracer of the dust-producing planetesimals, the various inclination 
determinations are compatible with coplanarity with the planetary orbits, 
and with the star, although the uncertainties remain large enough to preclude 
drawing a firm conclusion.

If the planetesimal belt and planet orbits are coplanar, as suggested by the 
derived inclination angle, then we would also expect alignment of the position 
angle of the belt and the argument of ascending node of the planet orbits, 
$\Omega$.  The \citet{konopacky16} astrometric analysis indicates $\Omega$ 
lies in the range of 40 to $70\degr$ for low eccentricity solutions, for all 
four of the planets. The disk position angle of $35\fdg6^{+9.4}_{-10.1}$ 
determined from the joint analysis of the SMA and ALMA data is compatible 
with the low end of this range. Like the inclination, the disk position angles 
derived from millimeter and far-infrared emission are discrepant, with the 
latter yielding a higher value of $64\degr\pm3\degr$ \citep{matthews14}.  
Again, there may be a physical origin for this difference. 
The overlap of the millimeter constraint on the 
planetesimal belt position angle and the planetary argument of ascending node 
provides weak additional support for coalignment of these components.

\subsection{Dynamical Constraints on the Mass of Planet b}
\label{sec:planetmass}
Because the wide-separation HR~8799 planets are relatively accessible to 
optical and near-infrared observations, they have become key benchmark objects 
for models of planetary atmospheres and evolution. Evolutionary models are 
highly degenerate in planet mass, luminosity and age, however, in part because 
the imprint of initial conditions can persist for 100's of Myr \citep{marley07}.
Depending on initial thermal content, giant planets can exhibit a wide range 
of luminosities. In principle, independent constraints on planet mass from
dynamics offer the potential to break these model degeneracies. Such 
constraints have the potential to help distinguish between planet formation 
models, in particular hot-start models where most entropy is retained, e.g. 
by gas that collapses directly to form planets, from cold-start models where 
most entropy is lost, e.g. by gas cooling in core accretion. Notably, the 
core accretion mechanism is challenged to form $5-10$~M$_{\rm Jup}$ planets 
{\em in situ} at the large orbital radii of the HR~8799 planets 
\citep{kratter10}. The observed luminosities of the HR~8799 planets cannot be 
reproduced by the most extreme cold-start models proposed by \citet{marley07}. 
However, they are compatible with the intermediate ``warm-start'' models 
discussed by \citet{spiegel12} and \citet{marleau14}, since luminosity 
increases with both mass and initial entropy (for a given age). These models
would imply much higher planet masses than hot-start models by up to a factor
of two (or perhaps more).  For planet b, a fully self-consistent solution 
for the planet properties remains elusive once spectroscopic constraints are 
combined with infrared photometry, likely due to limitations of the models 
\citep{marley12}. Dynamical constraints on planet masses would be valuable,
but they are generally difficult to obtain for wide-separation planets that 
are impractical to monitor over many orbital periods.  

For HR~8799, analysis of the long-term stability of the system offers one form 
of dynamical mass constraint on the planets.  The HR~8799 multi-planet system 
was investigated numerically by \citet{gotberg16} who
found that nominal models with 5--7~M$_{\rm Jup}$ 
planets are unstable on timescales much shorter than the system age.  
One natural solution to this problem is if the planet masses are 
substantially lower than estimated from infrared observations. 
Lower mass planets have larger separations in terms of Hill radii, and the 
system stability persists for much longer times.  If standard cooling models 
underpredict the planet luminosities, as some calculations suggest
\citep[e.g.][]{mordasini13}, then lower planet masses -- more like 
2--3 M$_{\rm Jup}$ -- might be accommodated.  Alternatively, much higher 
planet masses can be viable if the orbits are stabilized by resonant lock. 
The configuration of the HR~8799 planet orbits appear to be consistent 
with a 1:2:4:8 resonance \citep[e.g.][]{reidemeister09,konopacky16}.  

The truncation of the inner edge of the planetesimal disk by the gravity of 
the outermost planet~b offers another form of dynamical mass constraint. 
The separation of the planet from the disk is sensitive to the planet mass, 
as the planet clears a chaotic zone around itself due to resonance overlap
\citep[e.g.][]{quillen06a,chiang09}. 
For the simple case of circular and coplanar orbits, \citet{wisdom80} 
derived an analytic formula for the width of the chaotic zone exterior 
to the planetary orbit, $\Delta a = Ca_\text{pl}(M_\text{pl}/M_{*})^{2/7}$, 
where $C$ is a scaling coefficient of 1.3, $a_{pl}$ is the semi-major axis of 
the planet, and $M_{pl}$ and $M_{*}$ are the mass of the planet and the star, 
respectively. Other work that adopts a slightly different definition of 
resonance width gives a different scaling factor 
\citep[e.g. 1.4,][]{malhotra98}.
The width of the chaotic zone for more complex configurations has been the 
subject of numerous investigations, considering additional factors like the 
eccentricity of the planet orbit \citep{quillen06b,regaly17}, and 
eccentricities of the planetesimal orbits with the disk 
\citep{mustill12,pearce14}.  The \citet{pearce14} study formulates a
convenient expression for the location of the disk inner edge relative to the 
width of the chaotic zone, 
$R_\text{in} = a_\text{pl}+5a_\text{pl}(M_\text{pl}/3M_{*})^{1/3}$, taken in 
the limit of zero eccentricity.

\begin{figure}[h]
\begin{center}
\includegraphics[scale=1.00,angle=0]{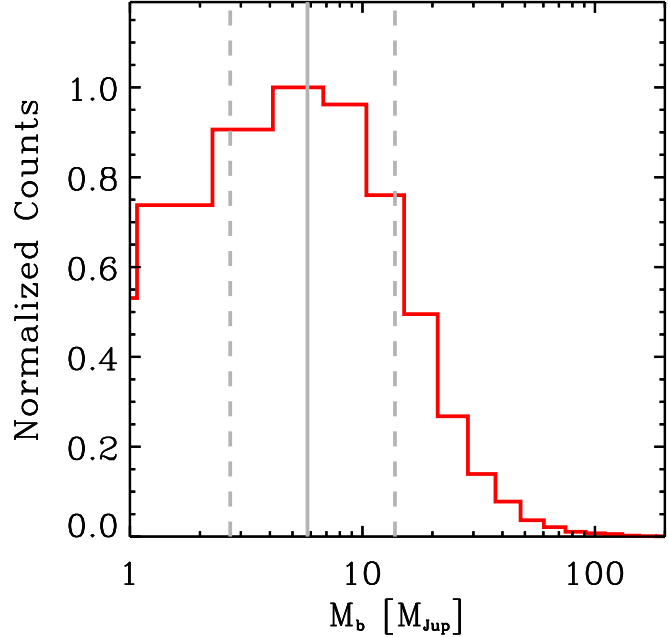}
\figcaption{
Constraint on the mass of planet~b obtained by assuming this planet
truncates the inner edge of the HR~8799 millimeter belt.  To generate this figure,
we have translated the inner radius posterior distribution from our 
MCMC modeling of the millimeter emission shown in Figure~\ref{fig:cornerplot}.
The solid gray line indicates the best-fit results, and the dashed gray lines
indicate the $1\sigma$ uncertainties.
}
\end{center}
\label{fig:planet_mass}
\end{figure}

If we adopt the \citet{pearce14} formula, then we can translate the 
radial location of the inner edge of the millimeter belt, 
$R_\text{in}= 104_{-12}^{+8}$~AU, directly into a constraint on the mass of 
the outermost planet~b. Figure~\ref{fig:planet_mass} shows the result 
of this translation of the inner radius posterior distribution from our 
modeling, assuming planet b truncates the belt.  
For this calculation, we adopt the planet semi-major axis $a_\text{pl}=68$~AU 
and stellar mass $M_* = 1.56$~M$_{\odot}$ from dynamical analysis 
\citep{soummer11}, 
consistent with optical photometry and spectral synthesis \citep{gray99}.
This procedure gives $M_\text{pl} = 5.8_{-3.1}^{+7.9}$~M$_{\rm Jup}$ for the 
planet b mass, compatible with standard planet cooling models, though the 
errors are still large enough to encompass both hot-start and a range of
warm-start 
initial conditions.  This new determination improves on previous attempts 
that use an inner belt radius estimated from the spectral energy distribution 
\citep{su09}, or from the ALMA data alone \citep{booth16}, both of which 
suffer from substantial systematic uncertainty.  Given this result, there is 
no need to posit the presence of an additional unseen planet orbiting beyond 
planet b, or variations in the orbit of planet b, to account for the disk 
truncation.  Since this planet mass constraint scales very steeply with 
the disk-planet separation, even a modest improvement in the 
constraint on the radial location of the inner edge of the millimeter belt 
has the potential to significantly reduce the uncertainty on the planet~b 
mass estimate.  Deeper observations of 
HR~8799 could be readily obtained with the SMA, or the Atacama Compact Array 
(ACA), at the appropriate spatial frequencies, to improve the constraint.
A better planet~b mass estimate obtained in this way could provide important 
feedback into the system dynamics, planet cooling models, and formation 
scenarios.

\subsection{Stirring Mechanisms}
\label{sec:stirring}
The planetesimals within debris disks like HR~8799 must be ``stirred'' in order
to incite destructive collisions \citep{wyatt08}. The two leading mechanisms
are self-stirring by an outwardly moving front of planetoid formation 
and growth \citep{kenyon02} and planet-stirring by the gravitational 
influence of interior giant planets \citep{mustill09}.  Given the young
age of HR~8799, \citet{moor15} concluded that the disk is most likely 
planet-stirred, since the long timescales needed to form Pluto-sized bodies 
at 100's of AU are not compatible with the large extent of the disk 
that is inferred from far-infrared imaging. The extent of the planetesimal 
belt more directly traced by millimeter emission supports this conclusion. 
Moreover, it seems likely that the giant planets responsible for stirring the 
debris disk are detected directly.

An additional observable that may help to discriminate between stirring 
scenarios is the radial distribution of planetesimals. Self-stirred models 
produce an extended planetesimal disk with an outwardly increasing surface 
density \citep[e.g., $\Sigma \propto r^{+7/3}$ in the models of][]{kennedy10}.
By contrast, standard steady-state debris disk models tend to produce surface 
density gradients that decrease with radius \citep[e.g.][]{krivov06}. 
High quality resolved millimeter observations of a handful of debris disks 
suggest either rising surface density profiles \citep{macgregor15} or hint at 
more complex radial structures 
\citep[e.g. multiple cold belts in HD 107146,][]{ricci15}.  From our model
fits for HR~8799, the best-fit power law index for the millimeter emission 
radial profile is $x=-0.38\pm0.47$. If we assume a radial temperature profile 
$T(r)\propto r^{-0.5}$ to approximate radiative equilibrium with stellar 
heating, then this implies a surface density profile 
$\Sigma(r)\propto r^{0.12\pm0.47}$.
While the uncertainties are still significant, this best-fit shallow 
dependence of planetesimal surface density on radius is not predicted in 
self-stirring scenarios. This result also favors planet-stirring as the 
basic collisional excitation mechanism for this radially extended debris disk. 

\subsection{Grain Size Distribution}
\label{sec:spectralindex}
The spectral index of dust emission at millimeter wavelengths from debris
disks encodes information on the grain size distribution that can be used 
to assess collisional models \citep{ricci12}.
For a power law grain size distribution, $n(a) \propto a^{-q}$, the 
reference model is the classical steady state collisional cascade where
$q=7/2$ \citep{dohanyi69}.  
Lower or higher values of $q$ may be obtained by varying the strength and
velocities of colliding bodies \citep{pan05,pan12}, and are found in 
time-dependent numerical models of debris disk evolution 
\cite[e.g.][]{schuppler15}.

For HR~8799, we use the best-fit total flux density from the combined SMA and 
ALMA models at 1.3~millimeters and the VLA flux density measured at 
9~millimeters
to obtain a millimeter spectral index of $\alpha_\text{mm} = 2.41\pm0.17$. 
Given a grain composition model, $\alpha_\text{mm}$ provides a constraint
on the size distribution of the emitting grains, in particular on the 
power-law index $q = (\alpha_\text{mm}-\alpha_\text{Pl})/\beta_s +3$.
In this expression, valid for $3<q<4$, $\beta_s$ is the dust opacity 
power law index in the small grain limit, $1.8\pm0.2$ for interstellar grain 
materials \citep{draine06}, and $\alpha_\text{Pl}$ is the spectral index 
of the Planck function, $1.92\pm0.10$ (the Rayleigh-Jeans limit modified by 
a small correction factor from the 39.7~K dust temperature assumed for 
HR~8799). Given the measured millimeter fluxes for HR~8799, the resulting 
grain size distribution power-law index is $q = 3.27\pm0.10$.  
This result is consistent with the weighted mean value of 
$q = 3.36\pm0.02$ determined by \citet{macgregor16} for a sample of 
15 debris disks observed over a similar wavelength range. This grain size
distribution is most consistent with a classical collisional cascade, 
with no need to invoke additional complexities in the collisional model.

\section{Conclusions}
\label{sec:conclusions}
We have made new observations of millimeter dust continuum emission from the 
debris disk that surrounds the four directly imaged planets in the HR~8799 
system.  SMA observations at 1.3~millimeters complement archival ALMA 
observations by adding measurements at lower spatial frequencies to better 
sample extended emission. By fitting simple parametric disk models to these 
data in an MCMC framework, we obtain constraints on the structural parameters 
of the millimeter emission. The main conclusions from this analysis are:

\begin{enumerate}
\item The 1.3~mm emission morphology imaged at $3\farcs8$ (150~AU) resolution 
is consistent with a broad ($\Delta R/R\gtrsim 1$), axisymmetric, 
inclined belt centered on the star.  These data provide no evidence for the 
presence of clumps or other azimuthal asymmetries that would betray dynamical 
sculpting of the disk by the gravity of the interior orbiting planets.

\item Model fits to the combined SMA and ALMA visibilities constrain the 
millimeter belt inclination, $32\fdg8_{-9.6}^{+5.6}$, and position angle, 
$35\fdg6_{-10.1}^{+9.4}$. These values overlap with estimates of the orbital 
inclination and angle of ascending node of the planet orbits determined by 
astrometric observations, and they are consistent with a coplanar configuration
of the disk and planetary orbits within the (still large) mutual uncertainties. 

\item The best-fit inner edge of the millimeter emission belt, 
$R_\text{in} =104_{-12}^{+8}$~AU, provides an independent dynamical constraint 
on the mass of the outermost planet~b of $5.8_{-3.1}^{+7.9}$~M$_{\rm Jup}$ 
under the assumption that the chaotic zone of this planet truncates the 
planetesimal disk. This dynamical mass estimate is commensurate with those 
obtained from the planet luminosity and standard hot-start evolutionary models,
although the uncertainties allow for a range of initial thermal content.

\item Flux density measurements at 1.3 and 9~millimeters from the SMA and VLA,
respectively, give a spectral index of $2.41\pm0.17$, which implies a grain 
size distribution power-law index of $q= 3.27\pm0.10$. This value is 
consistent with the weighted mean value of $q$ determined from millimeter 
observations of other debris disks and close to predictions for a steady state 
collisional cascade.

\end{enumerate}

The millimeter observations presented here have improved our view of 
the large scale spatial distribution of the dust-producing bodies in the 
debris disk surrounding the HR~8799 planetary system. However, these 
observations still lack the sensitivity needed to reveal any details 
of planet-disk interactions, or to obtain dynamical constraints on 
planet mass that usefully discriminate between planet evolutionary models. 
Deeper millimeter observations should be pursued to obtain better constraints 
on the planetesimal belt morphology, and to determine if more a sophisticated 
approach is needed for modeling and interpretation.

\acknowledgements
M.A.M. acknowledges support from the National Science Foundation under Award 
No. 1701406.
The Submillimeter Array is a joint project between the Smithsonian 
Astrophysical Observatory and the Academia Sinica Institute of Astronomy and 
Astrophysics and is funded by the Smithsonian Institution and the Academia 
Sinica.  
This paper makes use of the following ALMA data: ADS/JAO.ALMA\#2012.1.00482.S. 
ALMA is a partnership of ESO (representing its member states), NSF (USA) and 
NINS (Japan), together with NRC (Canada), MOST and ASIAA (Taiwan), and KASI 
(Republic of Korea), in cooperation with the Republic of Chile. The Joint ALMA 
Observatory is operated by ESO, AUI/NRAO and NAOJ. The National Radio Astronomy
Observatory is a facility of the National Science Foundation operated under 
cooperative agreement by Associated Universities, Inc.

\end{document}